\documentclass[12pt]{article}
\usepackage[a4paper,vmargin={30mm,30mm},hmargin={30mm,30mm}]{geometry}

\usepackage{amsmath}
\usepackage{amsfonts}
\usepackage{graphicx}
\usepackage{amsthm}
\usepackage[T1]{fontenc}
\usepackage{subfigure}
\usepackage[numbers]{natbib}

\usepackage{times}
\usepackage{url}

\newcommand{\sfigurecaption}{Four parameters were fitted: the basic reproduction number, $R_0$; the mean infectious period (equivalent to the generation time), $1/\gamma$; the reporting parameter, $r$; and the proportion
 of the population initially infectious. The orange line shows the assumed prior distribution for $1/\gamma$.}

\begin{document}

\subsection*{Modelling the transmission dynamics of online social contagion}

Adam J.~Kucharski$^1$ \\

\noindent $^1$Centre for the Mathematical Modelling of Infectious Diseases,\\ London School of Hygiene \& Tropical Medicine, London, UK \\ E-mail: adam.kucharski@lshtm.ac.uk

\subsection*{Abstract}

During 2014--15, there were several outbreaks of nominated-based online social contagion. These infections, which were transmitted from one individual to another via posts on social media, included games such as `neknomination', `ice bucket challenge', `no make up selfies', and Facebook users re-posting their first profile pictures. Fitting a mathematical model of infectious disease transmission to outbreaks of these four games in the United Kingdom, I estimated the basic reproduction number, $R_0$, and generation time of each infection. Median estimates for $R_0$ ranged from 1.9--2.5 across the four outbreaks, and the estimated generation times were between 1.0 and 2.0 days. Tests using out-of-sample data from Australia suggested that the model had reasonable predictive power, with $R^2$ values between 0.52--0.70 across the four Australian datasets. Further, the relatively low basic reproduction numbers for the infections suggests that only 48--60\% of index cases in nomination-based games may subsequently generate major outbreaks.

\subsection*{Introduction}

Since the start of 2014, there have been several large outbreaks of nomination-based social contagion on social networking sites such as Facebook and Twitter. Unlike much viral online content, which users typically share with their contacts in a spontaneous manner~\citep{bakshy2011everyone,cheng2014can,viralcharity}, the transmission of nomination-based content follows specific rules. Examples include the `neknomination' challenge~\citep{zonfrillo2014neknominate}, in which participants posted a video themselves finishing a drink online then nominated contacts to do the same, and the `ice bucket challenge', in which users posted footage of themselves getting doused with icy water~\citep{ni2014transmissibility}. Users have also nominated others to post `no makeup selfies'~\citep{deller2015selfies}, or to re-activate their first Facebook profile pictures~\citep{guardianFPP}.

The spread of social contagion can be studied using mathematical modelling frameworks developed for the analysis of infectious diseases~\citep{Hill:2010tg,House:2011mb}. In particular, it has been suggested that the population dynamics of the neknomination outbreak can be captured with a simple susceptible-infectious-recovered (SIR) model; based on the rules of the game, such a model predicted that the duration of outbreak would be less than a month~\citep{kucharski2014nom}. Nomination-based outbreaks have also been the subject of a retrospective cohort analysis: using individual-level transmission chains, it was possible to estimate the basic reproduction number (defined as the average number of secondary cases generated by a typical infectious individual in a fully susceptible population) and the serial interval of the ice bucket challenge~\citep{ni2014transmissibility}. 

As well as providing data with which to study the epidemiology of online contagion, nomination-based outbreaks can produce secondary benefits for the public health community, in the form of raised awareness of certain charities and causes. The ice bucket challenge came to be associated with the Amyotrophic Lateral Sclerosis (ALS) Association~\citep{koohy2014lesson}, and users posting no makeup selfies typically accompanied the picture with a donation to Cancer Research UK~\citep{deller2015selfies}. Understanding the dynamics of such outbreaks is therefore of interest to media teams and marketers as well as to epidemiologists and social network researchers.

Several aspects of nomination-based games remain little understood, however. In particular, it is unclear whether these outbreaks exhibit consistent epidemiological properties, and hence to what extent it is possible to predict such online contagion. Using a mathematical model, I examined the transmission dynamics of four nomination-based games that occurred during 2014--15. I estimated the reproduction number and generation time for each game in the United Kingdom, and tested the ability of the model to predict out-of-sample data from another country. I also compared the predicted duration of different outbreaks. Finally, I used the fitted model to examine the frequency of introduction of new infections, and hence estimate the proportion of new outbreaks that failed to take off.

\subsection*{Materials \& methods}

\subsubsection*{Data}

The analysis focused on four nomination-based online games. The first was the so-called `neknominate' (NKN) game, which emerged in early 2014~\citep{zonfrillo2014neknominate}. Each participant filmed themselves downing a drink, posted the video online, then typically nominated two or three of their contacts to do the same within 24 hours. The second nominated-based game started in March 2014, with users posting a `no makeup selfie' (NMS) photo, and choosing others to continue the chain. The third game was the `ice bucket challenge' (IBC), which appeared on social media during summer 2014. Users posted footage of themselves getting doused with icy water, and again nominated others to do the same. The fourth nomination-based outbreak started in January 2015, and involved users switching their Facebook profile picture (FPP) to the one they had when they first joined the network, and nominating contacts to follow suit.

I concentrated on the United Kingdom and Australia in the analysis because both countries experienced substantial transmission of all four games. As a proxy for disease incidence, I used Google Trends to assess interest level in the games over time~\citep{googletrends}. These data are based the number of searches for a particular term, which are then normalised  to generate a measure of daily `interest' between 0 and 100. Search terms included in the analysis were: `neknomination', `makeup selfie', `ice bucket challenge', `facebook first profile'. Google Trends data for the UK and Australia for the four outbreaks are shown in Figure~1.

\subsubsection*{Mathematical model}

Transmission in the population was modelled using a susceptible-infectious-recovered (SIR) framework~\citep{kermack115ag}. In the model, people who were initially susceptible became infectious when nominated by another person. Once they completed the task and nominated others, they recovered and could not become infectious again. The system of ordinary differential equations for the model was therefore as follows:
\begin{align}
dS/dt={}&-\beta SI \\
dI/dt={}&\beta SI -\gamma I \\
dR/dt={}&\gamma I \\
dC/dt={}&\beta SI 
\end{align}
where $S$ is the proportion of the population susceptible, $I$ is the proportion infectious, $R$ is the proportion that have recovered, and $C$ denotes the cumulative proportion infected in the model. In the model, $\beta$ is the transmission rate and $1/\gamma$ is the mean infectious period. Because people became infectious as soon as they performed the activity, the generation time of the disease---defined as the average time between an individual becoming infectious and infecting another person---was equal to the serial interval, the average time between the onset of symptoms (i.e.~doing the activity) in an infected host, and the onset of symptoms in the person they infect. The basic reproduction number was equal to $R_0=\beta/\gamma$.

To fit the model, I assumed the following observation process. Incidence on day $t$, denoted $c_t$, was defined as the difference in the cumulative proportion of cases over the previous day i.e. $c_t=C(t)-C(t-1)$. Hence $c_t$ represented the proportion of the population newly infected on day $t$. It was assumed the level of interest on day $t$ followed a Poisson distribution with mean $r c_t$, where $r$ represented the level of interest generated per 1\% of the population newly infected on a particular day. The earliest data point above 0 was used as the first observation date in the model fitting process, with the model initialised on the previous day.

In the model, a proportion $I_0$ of the population were initially infectious, and the rest of the population were susceptible. This resulted in four parameters to be estimated ($\beta$, $\gamma$, $r$, and $I_0$). Parameter estimation was performed using Markov chain Monte Carlo (MCMC). As the games often placed a 24 hour time limit on individuals to complete the activity~\citep{IBCrules,moss2015neknomination}, a gamma distribution was used as the prior for the mean infectious period, with $\mu=1$ day and $\sigma^2=0.1$. The posterior distribution was obtained from sampling over 40,000 MCMC iterations, after a burn-in period of 10,000 iterations (Figures~S1--S4). The model was implemented in R version 3.2.3~\citep{Rref}, with ODEs solved numerically using the \emph{ode45} method in the deSolve package~\citep{deSolvepackage}. 


\subsection*{Results}

Estimates for the basic reproduction number, $R_0$, ranged from 1.9 to 2.5 across the four outbreaks, with the highest estimate for NMS and lowest for the NKN and IBC games (Table~1). The estimated value of $R_0$ for the ice bucket challenge was 1.93 (95\% 1.63--2.36), which was higher than a previous study based on transmission chains generated by 99 well-known individuals~\citep{ni2014transmissibility}, in which $R_0$=1.43 (1.23--1.65). However, the estimate for mean generation time of 1.99 (1.46--2.85) days was consistent with the value of 2.1 days estimated in the earlier study~\citep{ni2014transmissibility}. Overall, the estimated generation time was largest for NKN and lowest for the FPP game. The fitted model suggested that the effective reproduction number, defined as $R(t)= R_0 S(t)$, dropped below the threshold value of one within a week or two of each outbreak starting (Figure~2).

The goodness of fit of the model was assessed using the coefficient of determination, $R^2$. Taking the maximum \emph{a posteriori} parameter estimates from the model for each outbreak and comparing simulated outbreaks using these parameters to the observed data produced $R^2$ values between 0.70--0.98 for the four outbreaks (Table~2), indicating that the model had reasonably good explanatory power. The predictive ability of the UK model was also tested, using out-of-sample data from Australia. This produced $R^2$ values between 0.52--0.70 (Table~2), suggesting that results from the simple UK model could to some extent be generalised to other countries.


I also explored the timescale of outbreak predicted by the model. Using 1000 samples from the posterior parameter estimates, I simulated outbreaks for each of the four nomination-based games starting with a fixed 1/1000 of the population infectious. For the NKN game, the outbreak peaked within 13.3 days ($\pm$~1.42~s.d.); for IBC it was 12.9 days ($\pm$~1.36), and 8.30 days ($\pm$~0.95) for NMS. The briefest predicted duration was FPP, which peaked in 4.60 days ($\pm$~0.52).

Finally, I estimated the probability that a new nominated-based game would fail to take off. When transmission is modelled as a standard branching process with Poisson offspring distribution and mean $R_0$, the probability that an outbreak starting with one initial case goes extinct is $1-1/R_0$~\citep{Lloyd-Smith:2005kl}. Hence an $R_0$=1.9--2.5 would suggest that only 48--60\% of index cases in nomination-based games successfully go on to generate a large outbreak. If transmission involves superspreading events (i.e.~there is individual-level variation in $R_0$) this proportion would be even smaller~\citep{Lloyd-Smith:2005kl}.

\subsection*{Discussion}

Using a simple SIR disease transmission model, I estimated  the basic reproduction number, $R_0$, and generation time of infection of four social contagion games. The estimates for $R_0$ were relatively consistent across outbreaks, with median values ranging from 1.5--2.5. For context, this range is similar to that estimated for acute infectious diseases such as the 2009 influenza A/H1N1p pandemic~\citep{fraser09} and the 2013-15 Ebola epidemic in West Africa~\citep{althaus2014estimating,nejm2014ebolaWHO}. The estimated generation time was shortest for the FPP outbreak, and longest for IBC and NKN, perhaps reflecting the ease with which each task could be performed.

There are several limitations to the analysis I have described. First I used Google Trends data as a proxy for disease incidence, which measures search interest in a particular topic, rather than the size of the outbreak itself. Other publicly available measures of interest, such as Twitter hashtags, would likely share this limitation, as mentioning a relevant keyword or hashtag is neither a necessary nor sufficient condition for participation in a nomination-based game. An alternative would be to analyse individual-level chains of transmission~\citep{ni2014transmissibility}. However, such data is likely to be extremely labour-intensive to collect: nominations are often made across different social media platforms, and in formats that are not easily machine-searchable (e.g. video nominations). In the absence of extensive individual-level data for the four outbreaks, I chose to use population-level data to fit the model. The resulting parameter estimates for IBC were of similar magnitude to parameters obtained from three generations of individual transmission chains, however, suggesting that this was a reasonable assumption.

I also assumed that the study population was fully susceptible initially, whereas there is evidence of variable susceptibility to social contagion~\citep{Aral:2012vn,moss2015neknomination}. However, if infectious individuals only nominate others who would be in the susceptible group at the start of the outbreak, then transmission should not be affected by the presence of non-initially susceptible individuals, as they will not form part of a potential nomination chain. In addition, I did not include potential for spatial synchrony as a result of nominations occurring across different regions. The second peak for NKN in Australia, which coincidences with the main outbreak in the UK (Figure 1), may be the result of such cross-continent nominations. Without detailed data on between-region transmission chains, however, it would be difficult to justify a more complex model, and I therefore chose to focus on a simple SIR framework in the analysis. 

To my knowledge, this is the first study to analyse outbreaks of nomination-based online contagion using a mechanistic mathematical model. The results demonstrate that simple epidemic models have the potential to be useful tool for examining the dynamics of such games. Moreover, the well-defined nature of these games means there is often information available about key parameters---such as generation time---which can be challenging to obtain for novel real-life pathogens. The relative stability of the parameter estimates across the four outbreaks also indicates that outbreaks of online nomination games may share fundamentally similar epidemiological properties. As well as providing a novel system on which to test epidemic modelling techniques, the extensive media and marketing interest in such outbreaks means that the insights gained from such models could also prove valuable in planning and predicting social media campaigns.



\footnotesize

\newpage

\begin{figure}[h]
  \centering 
   \includegraphics[width=0.7\textwidth]{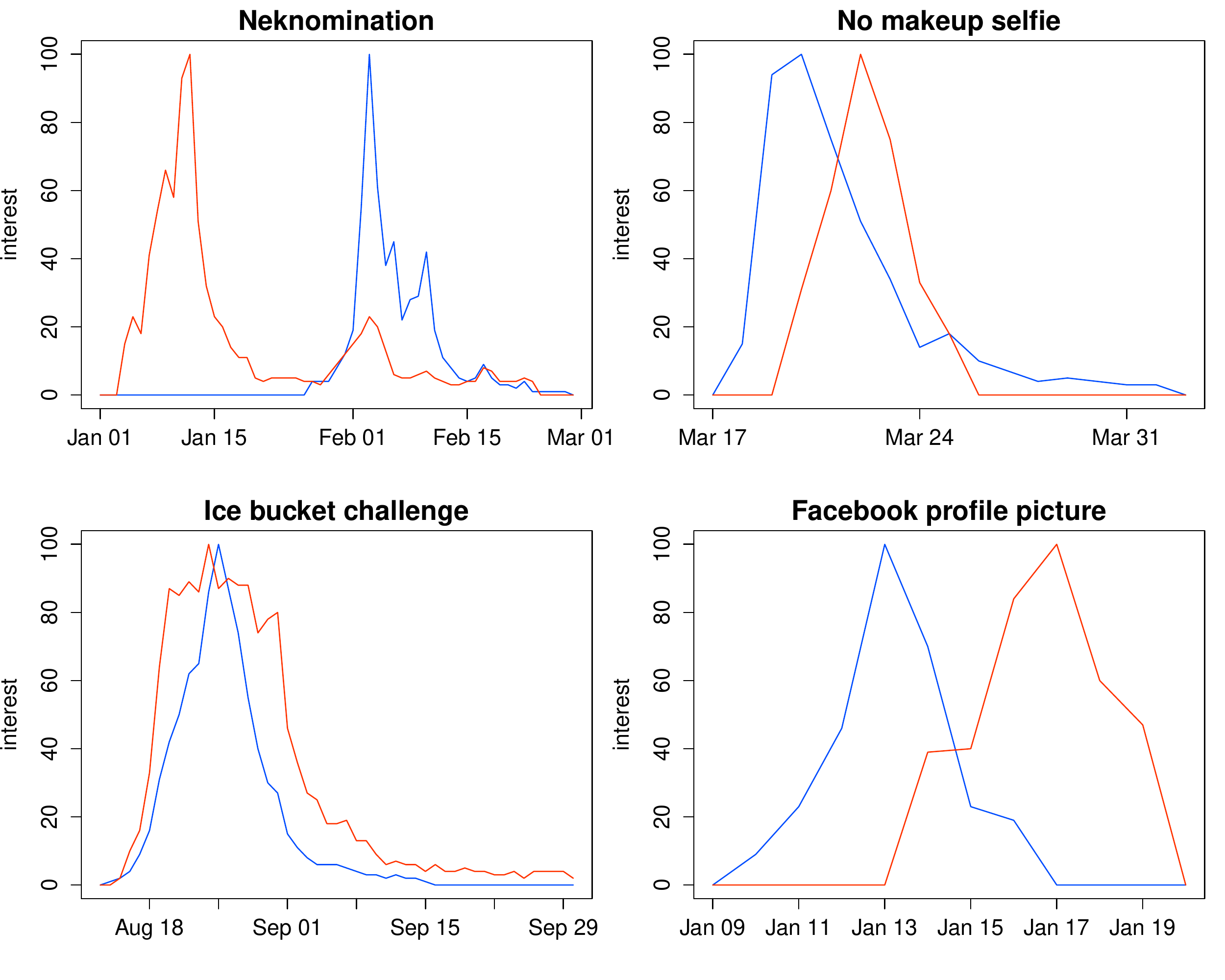} 
  \caption{Time series of Google search interest. Red lines, search terms in Australia; blue lines, interest in UK. Trends data are normalised so as to take values between 0--100. Data source: Google Trends (http://www.google.com/trends).}
  \label{fig:timeseries}
\end{figure}

\begin{figure}[h]
  \centering 
   \includegraphics[width=\textwidth]{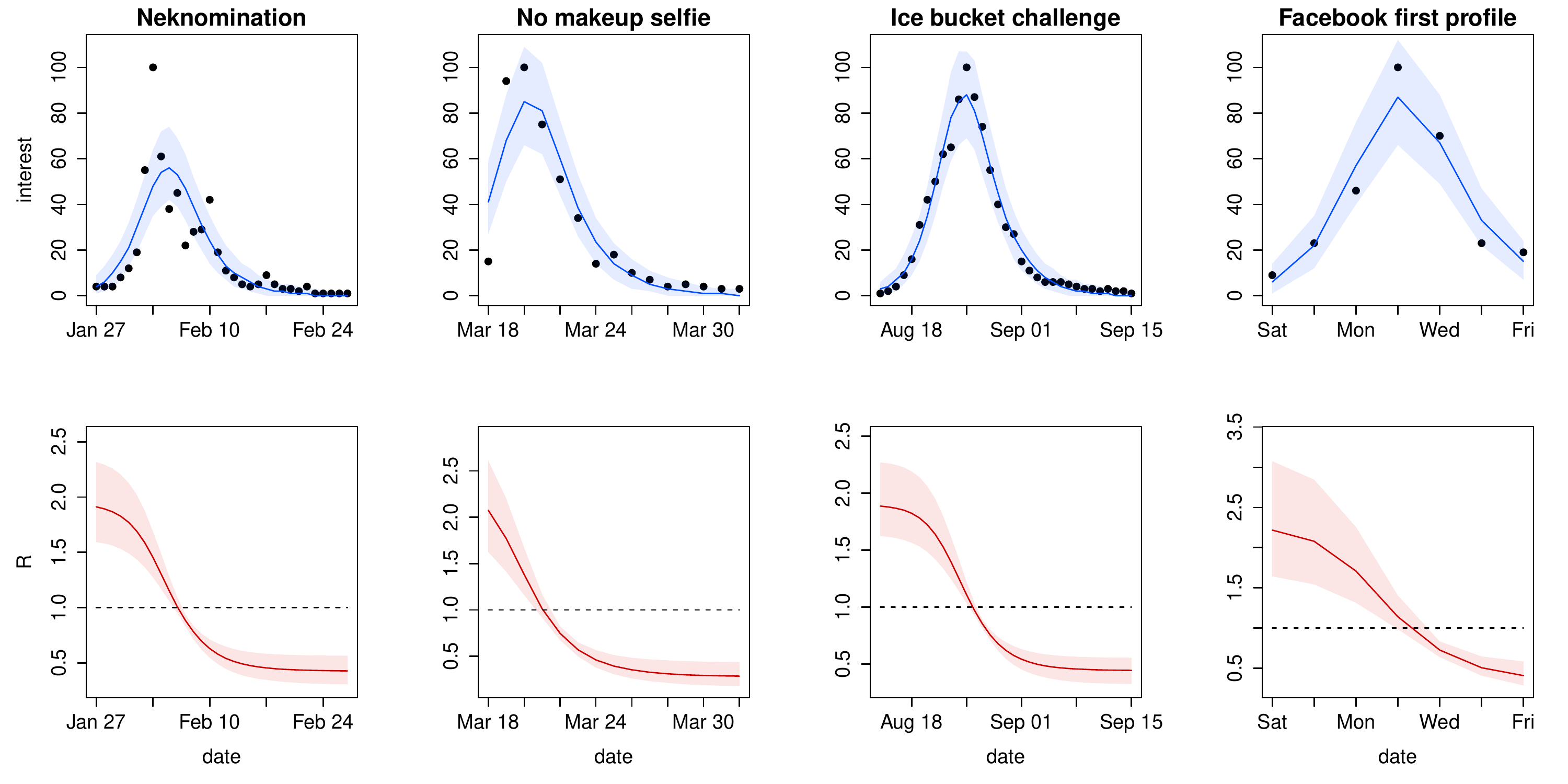} 
  \caption{Temporal evolution of different outbreaks of online social contagion in the UK. Top row: Model fit to outbreak time series. Black dots show Google Trends data, and blue lines show median interest levels from 1000 simulations of the fitted model, with parameters drawn from the joint posterior distribution; shaded region shows the 95\% credible interval. Bottom row: estimated change in effective reproduction number, $R=R_0 S$, over time. Lines shows median value from the 1000 simulations, with 95\% CI given by the shaded region.}
  \label{fig:modelfit}
\end{figure}

\newpage

\begin{table}[h]
\centering
\begin{tabular}{l l l}
\hline
Outbreak & $R_0$ & Generation time \\
\hline
Neknomination & 1.93 (1.57--2.35) & 2.04 (1.36--2.83) \\
No makeup selfie & 2.47 (1.92--3.14) & 1.70 (1.19--2.42) \\
Ice bucket challenge & 1.93 (1.63--2.36) & 1.99 (1.46--2.85) \\
Facebook profile picture & 2.29 (1.70--3.22) & 0.94 (0.56--1.61) \\
\hline
\end{tabular}
\caption{Posterior estimates for the basic reproduction number, $R_0$, and the generation time of infection. Point estimate gives the median of the distribution, with 95\% credible interval in parentheses.}
\label{tab:param}
\end{table}

\begin{table}[h]
\centering
\begin{tabular}{l l l}
\hline
Outbreak & UK & Australia \\
\hline
Neknomination & 0.702 & 0.515 \\
No makeup selfie & 0.889 & 0.699 \\
Ice bucket challenge & 0.980 & 0.678 \\
Facebook first profile & 0.935 & 0.579 \\
\hline
\end{tabular}
\caption{Goodness of model fit, as measured by the coefficient of determination, $R^2$. Model parameters were estimated from the UK data, and the maximum \emph{a posteriori} estimates were used to generate the model outputs tested against the two sets of observations.}
\label{tab:outsample}
\end{table}

\newpage~\newpage

\renewcommand{\thefigure}{S\arabic{figure}}
\setcounter{figure}{0}

\begin{figure}[h]
  \centering 
   \includegraphics[width=0.7\textwidth]{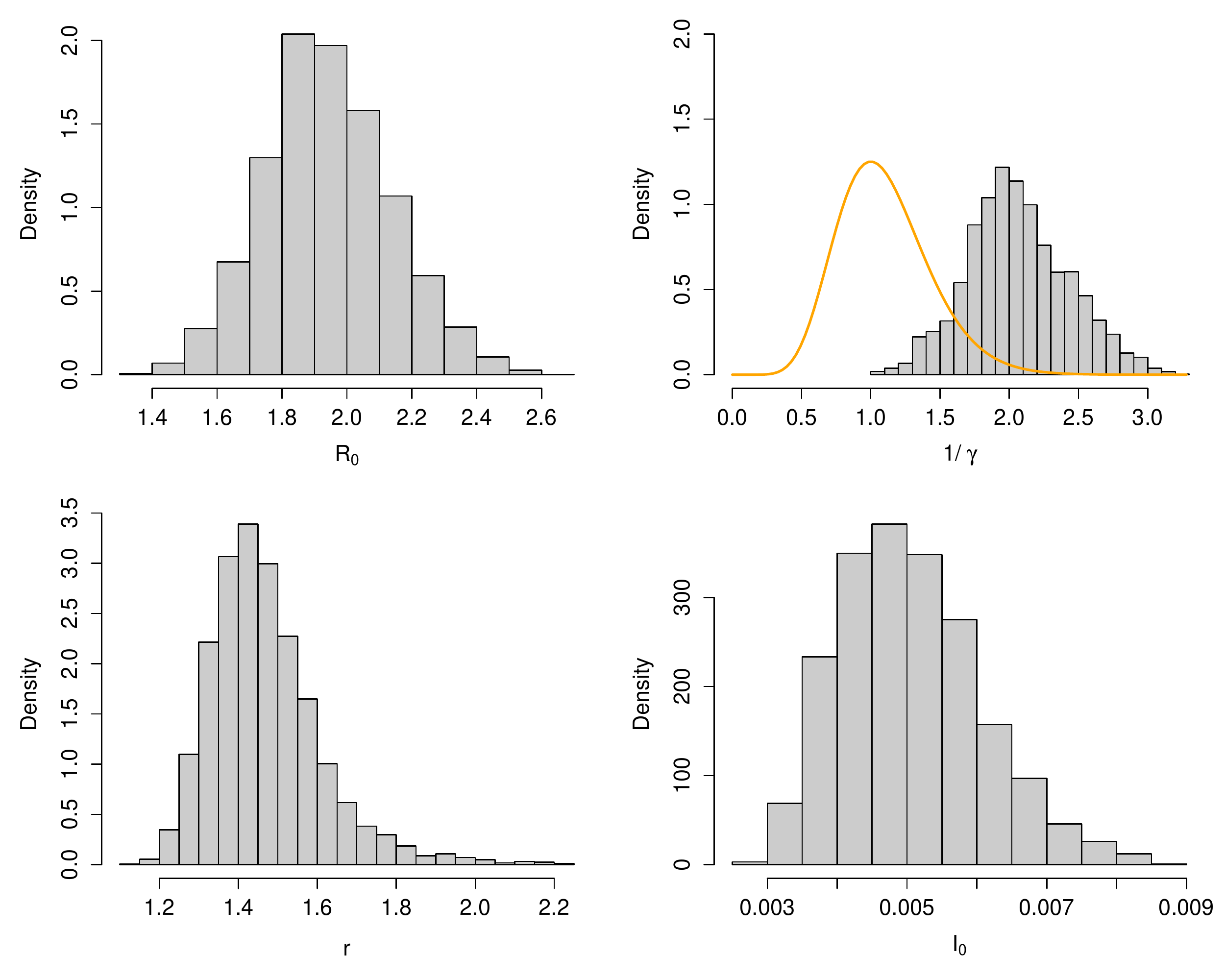} 
  \caption{Estimated posterior parameter distributions for the `neknomination' outbreak. \sfigurecaption}
  \label{fig:timeseries}
\end{figure}

\begin{figure}[h]
  \centering 
   \includegraphics[width=0.7\textwidth]{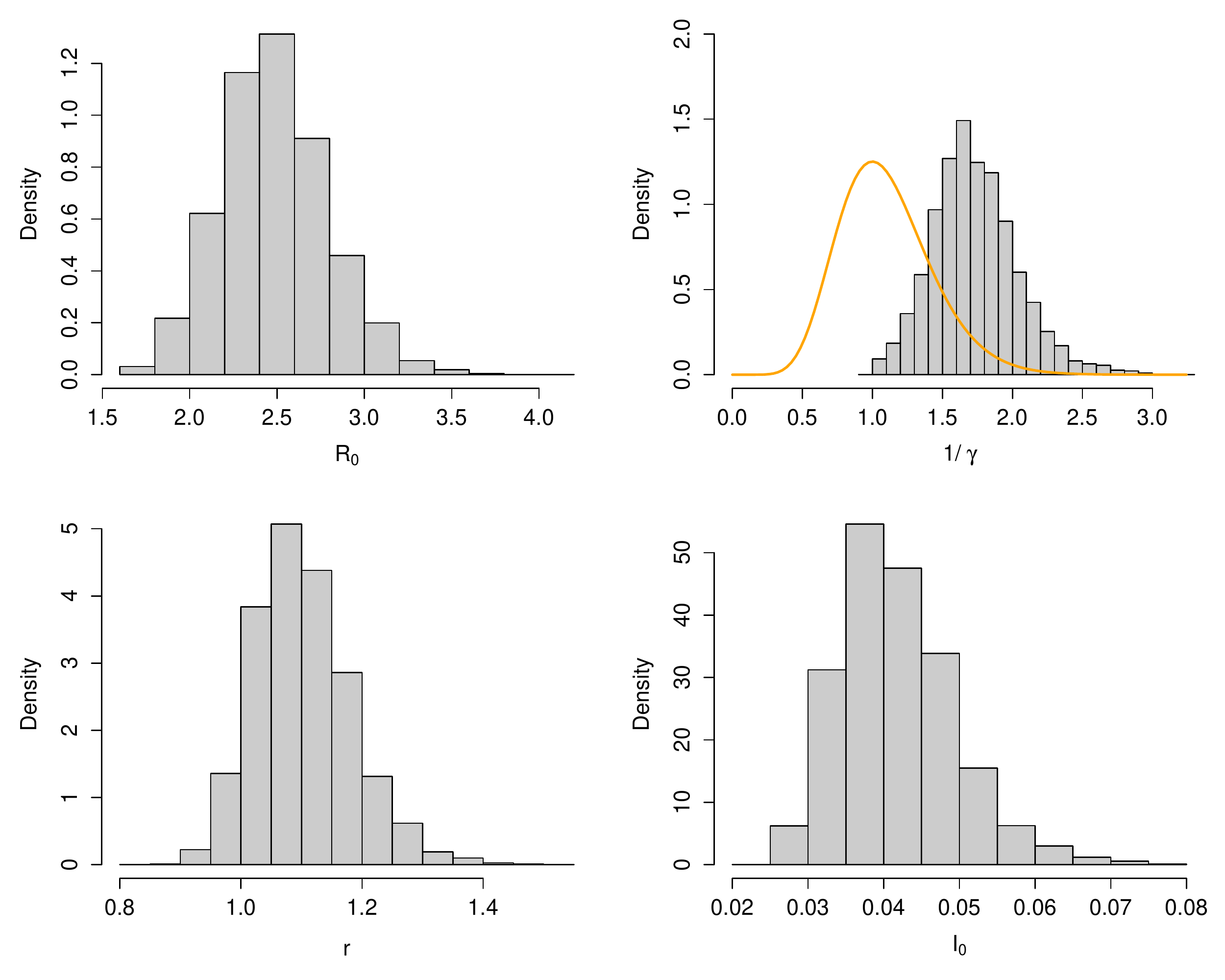} 
  \caption{Estimated posterior parameter distributions for the `no make up selfie' outbreak. \sfigurecaption}
  \label{fig:timeseries}
\end{figure}

\begin{figure}[h]
  \centering 
   \includegraphics[width=0.7\textwidth]{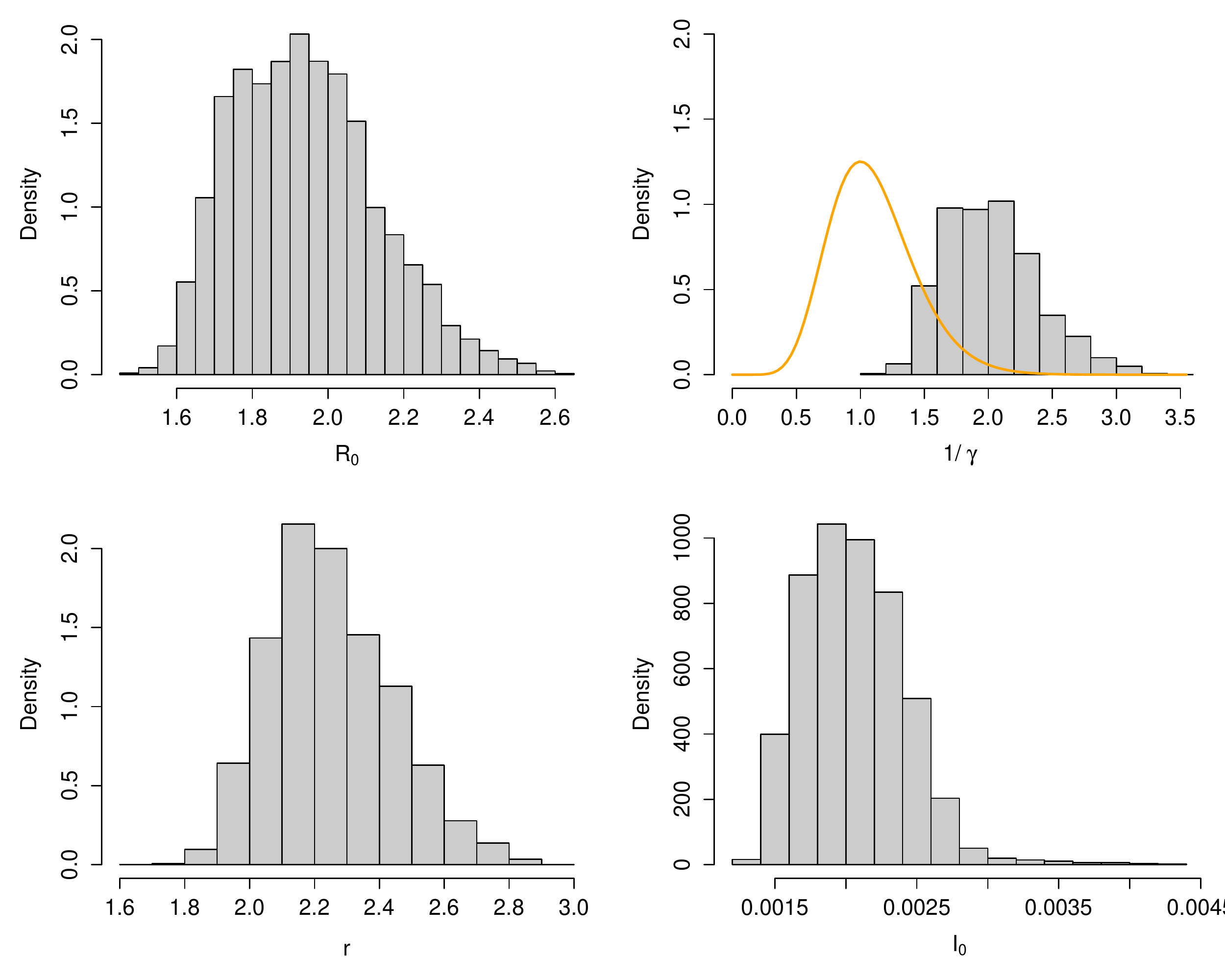} 
  \caption{Estimated posterior parameter distributions for the `ice bucket challenge' outbreak. \sfigurecaption}
  \label{fig:timeseries}
\end{figure}

\begin{figure}[h]
  \centering 
   \includegraphics[width=0.7\textwidth]{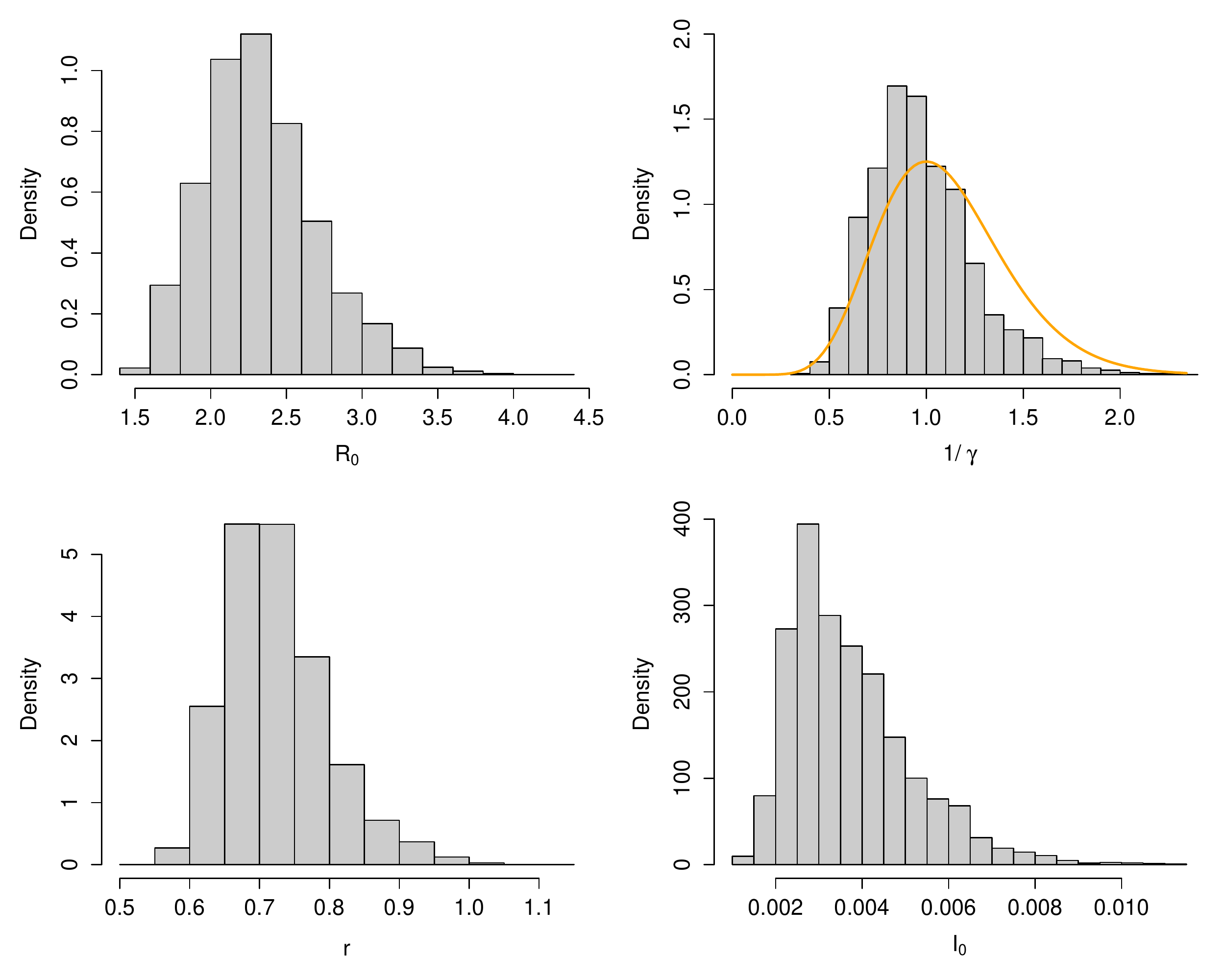} 
  \caption{Estimated posterior parameter distributions for the `Facebook profile picture' outbreak. \sfigurecaption}
  \label{fig:timeseries}
\end{figure}

\end{document}